\documentclass[aps,prl,twocolumn,groupedaddress,longbibliography]{revtex4-1}
\usepackage[dvipsnames]{xcolor}
\usepackage{graphicx,amsmath,amssymb,enumerate}

\begin{document}

\author{Robert S. Hoy}
\affiliation{Department of Physics, University of South Florida, Tampa, FL 33620 USA}
\email{rshoy@usf.edu}
\date{\today}

\title{Generating ultradense jammed ellipse packings using biased SWAP}

\begin{abstract}
Using a Lubachevsky-Stillinger-like growth algorithm combined with biased SWAP Monte Carlo and transient degrees of freedom, we generate ultradense disordered jammed ellipse packings.
For all aspect ratios $\alpha$, these packings exhibit significantly smaller intermediate-wavelength density fluctuations and greater local nematic order than their less-dense counterparts. 
The densest packings are disordered despite having packing fractions $\phi_{\rm J}(\alpha)$ that are within less than 0.5\% of that of the 
monodisperse-ellipse crystal [$\phi_{\rm xtal} = \pi/(2\sqrt{3}) \simeq .9069$] over the range $1.25 \lesssim \alpha \lesssim 1.4$ and coordination numbers $Z_{\rm J}(\alpha)$ that are within less than 0.5\% of isostaticity [$Z_{\rm iso} = 6$] over the range $1.3 \lesssim \alpha \lesssim 2.0$.
Lower-$\alpha$ packings are strongly fractionated and consist of polycrystals of intermediate-size particles, with the largest and smallest particles isolated at the grain boundaries.
Higher-$\alpha$ packings are also fractionated, but in a qualitatively-different fashion; they are composed of increasingly-large locally-nematic domains reminiscent of liquid glasses.
\end{abstract}
\maketitle

\section{Introduction}
\label{sec:intro}

Much attention has been paid over the past 20 years to jammed packings of anistropic particles and how they differ from those formed by disks and spheres \cite{donev04,delaney05,donev07,vanderwerf18,rocks23,mailman09,schreck10,schreck12,philipse96,williams03,desmond06,marschall18,jiao10,jiao11b,damasceno12,jaeger15,hoy17,maher21}. 
In parallel, over the past decade, the SWAP Monte Carlo algorithm \cite{grigera01,ninarello17} has enabled preparation of lower-$T$ equilibrated supercooled liquids, more-stable glasses, and denser disordered jammed packings than was previously feasible \cite{berthier16,berthier16b,ozawa17,ozawa18,berthier19d,scalliet19,scalliet22}.
Recent work has shown that allowing particles' diameters to vary independently during sample preparation provides additional \textit{transient} 
degrees of freedom (TDOF) which can be exploited to obtain even-stabler glasses and even-denser packings \cite{kapteijns19,hagh22,bolton24}.

Surprisingly, however, the latter two developments have not yet been exploited to shed light on the first topic.
More generally, very few simulation studies have attempted to determine how jammed anistropic-particle packings' structure depends on their preparation protocol, despite the great insights obtained from comparable studies of disk and sphere packings \cite{torquato00,donev04c,chaudhuri10,ozawa17} and the many open science questions raised by recent experimental studies of anistropic-particle (colloidal and small-molecule) glasses with strongly-preparation-protocol-dependent multiscale structure \cite{zheng11,mishra13,roller20,roller21,liu15b,liu17b,gujral17,teerakapibal18}.

This combination of factors presents an opportunity to make progress on multiple fronts by applying SWAP and TDOF moves during the preparation of jammed anistropic-particle packings.
Two-dimensional ellipses are perhaps the best shapes with which to begin such an effort, since they are a straightforward generalization of disks and their jamming phenomemology for preparation protocols which mimic \textit{fast} compression has already been extensively studied \cite{delaney05,donev07,vanderwerf18,rocks23,mailman09,schreck10,schreck12}.
In this paper, we show that adding a suitably \textit{biased} SWAP algorithm and a minimalistic implementation of TDOF to a Lubachevsky-Stillinger (LS)-like particle-growth algorithm \cite{lubachevsky91} yields jammed ellipse packings which are significantly denser than any previously reported for all $1 < \alpha \leq 5$.
These packings' multiscale structure differs qualitatively from that of their less-dense counterparts, in a nontrivial and strongly-$\alpha$-dependent fashion.

\section{Methods}
\label{sec:methodsl}

We recently performed a detailed characterization of jammed ellipse packings' structure \cite{rocks23} over a much wider range of aspect ratios ($1 \leq \alpha \leq 10$) than had been considered in previous studies \cite{delaney05,donev07,vanderwerf18,mailman09,schreck10,schreck12}.
To understand the effects of particle dispersity, we employed three different probability distributions for the ellipses' inital minor-axis lengths $\sigma$:
\begin{equation} 
\begin{array}{c}
P_{\rm mono}(\sigma) = \delta(\sigma - .07) , \\
\\
P_{\rm bi}(\sigma) = \displaystyle\frac{\delta(\sigma - .05 a)}{2} + \displaystyle\frac{\delta(\sigma - .07)}{2} ,\\
\\
\textrm{and}\ \ P_{\rm contin}(\sigma) =  \Bigg{ \{ }  \begin{array}{ccc}
\displaystyle\frac{7}{4\sigma^2} & , & .05 \leq \sigma \leq .07\\
\\
0 & , & \sigma < .05\ \textrm{or}\ \sigma > .07
\end{array} ,
\end{array} 
\label{eq:dispersity} 
\end{equation}
where $\delta$ is the Dirac delta function and {$\sigma$ is expressed in arbitrary units of length.}
For all but the smallest aspect ratios (where systems with $P = P_{\rm mono}$ formed jammed states with a high degree of crystallinity, as expected \cite{lubachevsky91}), all three of these $P(\sigma)$ produced the same \textit{qualitative} structural trends.
For example, the densest jammed packings always had the best-ordered first coordination shells, exhibiting positional-orientation correlations which were substantially greater than those of their less-dense counterparts, even though the details of these correlations were strongly $P(\sigma)$-dependent.

Choosing $P = P_{\rm contin}$ produces systems in which equal areas are occupied by particles of different sizes, and apparently optimizes glass-formability for a wide variety of interparticle force laws \cite{ninarello17}.
Moreover, in contrast to $P_{\rm bi}$, which has been employed as the standard model for granular materials over the past 20 years \cite{ohern03} and 
was the only $P(\sigma)$ employed in all other previous studies of ellipse jamming \cite{donev07,delaney05,vanderwerf18,mailman09,schreck10,schreck12}, it
allows for efficient particle-diameter swapping \cite{ninarello17}.

We made no attempt in Ref.\ \cite{rocks23}, however, to employ SWAP or indeed to investigate preparation-protocol dependence in any way.
Instead, all packings were generated using the same protocol: a LS-like particle-growth algorithm \cite{lubachevsky91} that mimicked \textit{rapid} compression. 
Each growth cycle consisted of two steps:
\begin{enumerate}
\item Attempting to translate each particle $i$ by a random displacement along each Cartesian direction and rotate it  by a random angle; and
\item Increasing \textit{all} particles' minor-axis lengths $\sigma$ by the \textit{same} factor $\tilde{\mathcal{G}}$, where $\tilde{\mathcal{G}}$ is the value that brings \textit{one} pair of ellipses into tangential contact.
\end{enumerate}
Here we obtain substantially higher jamming densities by adding two more steps to this cycle:
\begin{enumerate}
\setcounter{enumi}{2}
\item SWAP moves which exchange the diameters of larger particles  with smaller ``gaps'' (defined below) with those of smaller particles with larger gaps; and
\item TDOF moves which grow particles by \textit{different} factors $\mathcal{G}_i$ and thus allow the \textit{shape} of $P(\sigma)$ to vary.
\end{enumerate}

As in Ref.\ \cite{rocks23}, we begin by placing $N = 1000$ nonoverlapping ellipses of aspect ratio $\alpha$, with random positions and orientations, and minor-axis-length distributions given by $P = P_{\rm contin}$, in square $L \times L$ domains with $L = \sqrt{N\alpha}$.
Periodic boundary conditions are applied along both directions, so these initial states have packing fractions $\phi < 0.01$.
Then we begin the particle-growth procedure, which executes steps 1-3 for each growth cycle throughout the run, and step 4 in the latter stages of the run.
Overlaps between ellipse pairs $(i,j)$ are prevented throughout the entire process using Zheng and Palffy-Muhoray's exact expression \cite{zheng07} for their orientation-dependent distance of closest approach $d_{\rm cap}(i,j)$.

In step (1), the attempted translations and rotations have maximum magnitudes $0.05f$ and $(16f/\alpha)^\circ$, respectively. 
The move-size factor $f$ is set to $1$ at the beginning of all runs, and multiplied by $3/4$ whenever $100$ consecutive growth cycles have passed with $\tilde{\mathcal{G}} < 10^{-10}$.
Runs are terminated and the configurations are considered jammed when $f$ drops below $2\times 10^{-8}$.
These cutoff values for $f$ and $\tilde{\mathcal{G}}$ are the smallest values allowed by our double-precision numerical implementation.

In step (2), the fractional particle-growth rate per cycle is set to the maximum value which does not introduce any interparticle overlaps, i.e. by $\tilde{\mathcal{G}} = \min(\mathcal{G}_i)$, where
\begin{equation}
\mathcal{G}_i = \min\left[ \displaystyle\frac{\sigma_i}{2\alpha (\sigma_i + \sigma_j)} g_{ij} \right].
\label{eq:fracgrowthrate}
\end{equation}
The gap distances $g_{ij}$ are defined using the relation $g_{ij} = r_{ij} - d_{\rm cap}(i,j)$, so the quantity within the square brackets is a \textit{lower} bound for the amount by which particles $i$ and $j$ can grow without overlapping: specfically, it is the factor by which particles $i$ and $j$ can grow without overlapping if they are aligned \textit{end-to-end}.
The minimum in Eq.\ \ref{eq:fracgrowthrate} is taken over all nearest neighbors ($j$) of particle $i$, while the subsequent minimum defining $\tilde{\mathcal{G}}$ is taken over all $i$.
These choices make the algorithm more efficient by allowing particles to grow slower when gaps are small and faster when they are large.
We emphasize that imposing a uniform growth rate $\tilde{\mathcal{G}}$ preserves the \textit{shape} of the particle-size distribution $P(\sigma)$ defined in Eq.\ \ref{eq:dispersity}.
In other words, the ratio $\sigma_{\rm max}/\sigma_{\rm min} = 1.4$ of the largest and smallest ellipses' minor-axis lengths, and indeed the ratios of all other moments of $P(\sigma)$, remain constant as $\langle \sigma \rangle = \int_{\sigma_{\rm min}}^{\sigma_{\rm max}} \sigma P(\sigma) d\sigma$ increases.

Step (3) begins by recalculating all the $g_{ij}$ and then re-indexing particles in order of increasing $\tilde{g}_i = \min(g_{ij})$, where the minimum is again taken over particle $i$'s nearest neighbors.
Then, for each $i < N$, a particle index $k > i$ is randomly selected; the corresponding particles necessarily have $\tilde{g}_k > \tilde{g}_i$.
If they also have $\sigma_k < \sigma_i$, the SWAP move is accepted, the $\tilde{g}$ values for particles $i$, $k$ and their nearest neighbors are recalculated to reflect the new configuration, and the re-indexing is repeated.
If, on the other hand, the selected particle has  $\sigma_k > \sigma_i$, the move attempt is canceled and a different $k$-value (i.e., a different potential SWAP partner) is selected.
When either a swap has been completed or $N/10$ $k$-values have been sampled without finding a particle with $\sigma_k > \sigma_i$, the algorithm proceeds to the next particle (the next $i$ value).
This procedure yields high SWAP-move success rates, particularly when $\phi$ is still low.\footnote[1]{Success rates only become small when either the ordering of the $\tilde{g}_i$ amongst the $N$ particles parallels the ordering of their $\sigma_i$, or when most of the $\tilde{g}_i$ drop to zero.}

Step (4) also begins by recalculating all the $\tilde{g}_i$ and then re-indexing particles in order of increasing $\tilde{g}_i$.
Then it proceeds by growing each particle by a factor $\min(\mathcal{G}_i, 10^{-3})$; this cap on the growth rate prevents particles with unusually large $\tilde{g}_i$ from growing too quickly.
In contrast to step (2), step (4) allows $P(\sigma)$ to vary, and effectively adds one \textit{transient} DOF per particle \cite{kapteijns19,hagh22,bolton24}.
Note that this step is executed only if $f < 10^{-2}$.
We found that this choice both maximizes the final $\phi_{\rm J}(\alpha)$ and keeps increases in systems' polydispersity over the course of the packing-generation runs very modest.

Critically, in contrast to standard hard-particle SWAP \cite{grigera01} which accepts any move that does not introduce interparticle overlap, our procedure is \textit{biased} towards both increasing the minimum value of $\tilde{g}_i$.
By effectively introducing an ``energy'' cost for nonuniform $\{ \tilde{g}_i \}$, both the SWAP moves and the TDOF moves  act in a similar spirit to the TDOF moves employed in Refs.\ \cite{kapteijns19,hagh22,bolton24}.  
Specifically, they both decrease the width of the probability distributions $P(\tilde{g})$ by systematically transferring mass from regions with smaller gaps to regions with larger gaps.
The SWAP moves accomplish this while leaving the packing fraction unchanged, while the TDOF moves produce a spatially-nonuniform densification rate.

\section{Results}
\label{sec:results}

In this section, we will both qualitatively and quantitatively compare the structure of jammed ellipse packings generated using different sample-preparation protocols.
Novel results obtained using all four steps of the growth algorithm described above were averaged over 25 independently prepared samples.
Results obtained using only steps (1-2) of this algorithm are taken from Ref.\ \cite{rocks23}.
Ref.\ \cite{delaney05}'s were generated using a LS-like algorithm similar in spirit to (if different in its details from) that detailed in steps (1-2).
Ref.\ \cite{donev07}'s were obtained using the standard LS algorithm \cite{lubachevsky91,donev05}.
Ref.\ \cite{vanderwerf18}'s  and Refs.\  \cite{mailman09,schreck10,schreck12}'s were obtained by successive cycles of compression followed by conjugate gradient (CG) energy minimization; their $\phi_{\rm J}(\alpha)$ were identified as the packing fractions above which potential energy no longer dropped to zero.
In some figures, we will show data from refs.\ \cite{delaney05,donev07,vanderwerf18} to illustrate the variety of results obtained in previous studies of ellipse jamming.
Results from Refs.\ \cite{mailman09,schreck10,schreck12} followed the same general trends, and will be omitted for clarity.

\begin{figure}[!htbp]
\includegraphics[width=3in]{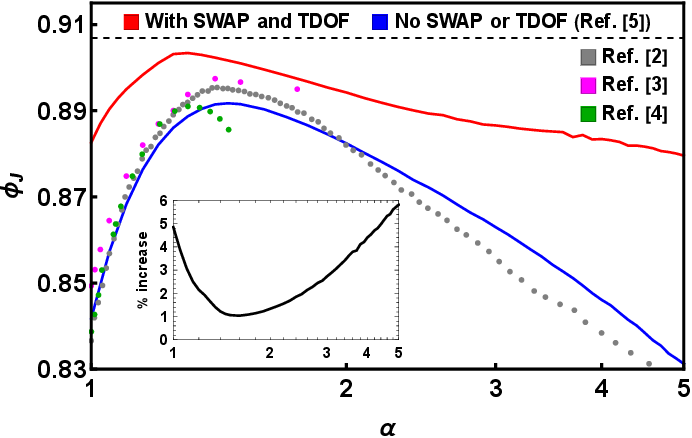}
\caption{ Jamming densities of systems prepared with and without SWAP and TDOF moves.  
The dashed line indicates $\phi_{\rm xtal} \simeq .9069$, and the inset shows the percentage increases over the $\phi_J(\alpha)$ obtained in Ref.\ \cite{rocks23} obtained by adding steps (3-4) to the particle-growth procedure.}
\label{fig:1}
\end{figure}

Figure \ref{fig:1} shows the preparation-protocol dependence of $\phi_{\rm J}(\alpha)$.
Adding SWAP and TDOF moves always generates substantially denser packings, but the degree to which this is so, and the structural differences associated with the density improvement, are strongly $\alpha$-dependent.
The packing fraction obtained for disks, $\phi_{\rm J}(1) \simeq .883$, is consistent with previous studies of collectively jammed monodisperse disk packings \cite{donev04c}, which are typically highly crystalline.
Polydisperse disk packings with such high densities were not reported until very recently.
Refs.\ \cite{bolton24,kim24} used  sophisticated SWAP and/or TDOF-based algorithms to obtain even denser packings, which had  $.89 \lesssim \phi_{\rm J} \lesssim .91$  despite remaining amorphous, but the methods employed in these studies are not readily generalizable to anistropic particles.

The packing-efficiency gain from adding SWAP and TDOF moves decreases monotonically from $\sim\!\! 5\%$ to $\sim\!\! 1\%$ as $\alpha$ increases from $1$ to $1.6$.
This rapid decrease makes the shape of the $\phi_J(\alpha)$ curve obtained using SWAP and TDOF moves differ in two key ways from those obtained without these moves, including results from previous studies \cite{delaney05,donev07,vanderwerf18,rocks23}.
First, the initial slope $(\partial\phi_{\rm J}/\partial \alpha)_{\alpha = 1}$, whose positive value demonstrates that anisotropic particles' ability to rotate away from one another allows them to pack more densely than disks \cite{donev04,delaney05,donev07}, is much smaller when SWAP and TDOF moves are employed, suggesting that the density-enhancing effect of allowing particle rotations weakens as systems get denser.

Second, the aspect ratio $\alpha_{\rm max}$ at which $\phi_{\rm J}(\alpha)$ is maximized gets shifted to lower values.
Specifically, while Refs.\  \cite{delaney05}, \cite{donev07} and \cite{rocks23} respectively found $\alpha_{\rm max} = 1.43$, $\alpha_{\rm max} = 1.40$ and $\alpha_{\rm max} = 1.45$, here we find $\alpha_{\rm max} = 1.30$.
Ref.\ \cite{vanderwerf18} also found $\alpha_{\rm max} = 1.30$; the fact that this result was similar to ours, rather than those from Refs.\ \cite{delaney05,donev07,rocks23}, probably owes to the authors' choice of sample-preparation protocol.
CG minimization of dense systems generates forces which can transmit stress over substantial distances, and hence (much like biased-SWAP and TDOF moves) tend to suppress long-wavelength density fluctuations.

Refs.\ \cite{delaney05,donev07,vanderwerf18,rocks23} respectively found $\phi_{\rm J}(\alpha_{\rm max}) = .895$, $.8974$, $.891$ and $.8917$.
Here we find $\phi_{\rm J}(\alpha_{\rm max}) = .9034$, which
is less than 0.4\% below $\phi_{\rm xtal}$.
Although this packing fraction is only $\sim\! \! 1\%$ larger than the largest value reported in previous studies of ellipse jamming, it reduces the minimum values of the void area fractions $\phi_{\rm v}(\alpha) = \phi_{\rm xtal} - \phi_{\rm J}(\alpha)$ by 71\%, 64\%, 79\%, and 78\% from those reported in Refs.\ 
\cite{delaney05,donev07,vanderwerf18,rocks23}, respectively.
In other words, the densest packings we obtain using SWAP and TDOF moves have \textit{far} less ``free volume'' than those obtained in previous studies.
Comparably large reductions in free volume persist over a wide range of $\alpha$.
For example, we find that $\phi_{\rm J}(\alpha) > .995\phi_{\rm xtal}$ [and hence $\phi_{\rm v}(\alpha) < .005\phi_{\rm xtal}$] for all $1.25 \lesssim \alpha \lesssim 1.40$.
Here we have implicitly assumed that $\phi_{\rm xtal}$ is the maximum possible packing fraction.
This hypothesis has been proven correct for monodisperse ellipses \cite{toth50}, and no denser polydisperse ellipse packings have been reported to the best of our knowledge.
On the other hand, Ref.\ \cite{bolton24} found $\phi_{\rm J}(1) > \phi_{\rm xtal}$ in systems with a substantially larger polydispersity index than those considered here, and a more advanced algorithm might be able to achieve the same result for $\alpha > 1$.

For $\alpha > 1.6$, the packing-efficiency gain increases monotonically, reaching $\sim\!\! 6\%$ by $\alpha = 5$.
This rapid increase causes the shape of the $\phi_J(\alpha)$ curve to differ in a third key way from those reported in previous studies.
Specifically, the rapid decrease of $\phi_{\rm J}(\alpha)$ for $\alpha > 2$  \cite{delaney05,vanderwerf18,rocks23},
which  is widely believed to be a general feature of anisotropic-particle jamming  \cite{philipse96,desmond06} provided systems remain isotropic as they are compressed, is sufficiently strongly suppressed that $\partial^2 [\ln(\phi_{\rm J})]/\partial [\ln(\alpha)]^2$ is positive rather than negative.
In other words, the slow crossover to $\phi_{\rm J} \sim 1/\alpha$ scaling expected from Onsager-like arguments \cite{onsager49} and evident in the $\phi_{\rm J}(\alpha)$ curves presented in Refs.\ \cite{delaney05,rocks23} is absent when SWAP and TDOF moves are employed, at least for the range of $\alpha$ considered here.
Below, we will argue that this qualitative difference is made possible by the moves' tendency to increase packings' orientational order.

Previous work on ellipse jamming has devoted much attention to $Z_{\rm J}(\alpha)$ because it illlustrates several key features of how anisotropic particles pack.
Since smooth 2D convex anisotropic particles have three degrees of freedom (two translational, one rotational), one would naively expect them to jam at isostaticity ($Z_{\rm J} = Z_{\rm iso} = 6$).
This behavior, however, has not been observed in previous studies of ellipses  \cite{delaney05,donev07,vanderwerf18,rocks23,mailman09,schreck10,schreck12}, spherocylinders \cite{vanderwerf18,marschall18}, or superdisks \cite{jiao10}.
Instead, all previous studies of ellipses have found a square-root singularity at small aspect ratios [$Z_{\rm J} - 4 \propto \sqrt{\alpha - 1}$ for $\alpha - 1 \ll 1$]
and a substantially-hypostatic plateau at intermediate aspect ratios [$5.5 \lesssim Z_{\rm J} \lesssim 5.8$ for $1.5 \lesssim \alpha \lesssim 3$].
These trends have been interpreted in terms of particles being mechanically stabilized by their curvature at the point of contact \cite{donev07} and/or by  \textit{quartic} vibrational modes \cite{mailman09,schreck10,schreck12}, but in light of the protocol-dependence of $\phi_{\rm J}(\alpha)$ discussed above, it is worth revisiting the protocol-dependence of $Z_{\rm J}(\alpha)$ here.

\begin{figure}[!htbp]
\includegraphics[width=3in]{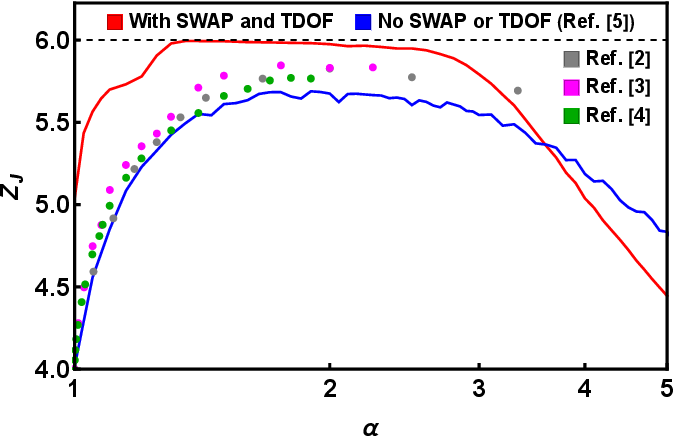}
\caption{
Coordination numbers of systems prepared with and without SWAP and TDOF moves.  The dotted line indicates $Z_{\rm iso} = 6$.}
\label{fig:2}
\end{figure}

Figure \ref{fig:2} shows that adding SWAP and TDOF moves increases $Z_{\rm J}$ by $\sim 1$ for small aspect ratios, e.g.\ from $4.02$ to $5.04$ for $\alpha = 1$.
After going through a minimum in $\partial Z_{\rm J}/\partial \alpha$ at $\alpha = 1.1$ which will be discussed further below, the coordination numbers again increase rapidly until reaching a plateau.
Systems have $Z_{\rm J} > .995 Z_{\rm iso}$ over a very wide range of aspect ratios ($1.3 \lesssim \alpha \lesssim 2.0$), and over a narrower range of $\alpha \gtrsim \alpha_{\rm max}$ (specifically, $1.35 \lesssim \alpha \lesssim 1.55$), they have $Z_{\rm J} > .998Z_{\rm iso}$.
These values were calculated without attempting to remove ``rattlers.''
Much as the results shown in Fig.\ \ref{fig:1} indicated a dramatic decrease in the free volume $\phi_{\rm v}(\alpha)$ despite the relatively modest absolute increases  in $\phi_{\rm J}(\alpha)$, those reported in Fig.\ \ref{fig:2} (at least for $\alpha \lesssim 3$) indicate an even more dramatic decrease in the degree of hypostaticity $H(\alpha) = Z_{\rm iso} - Z_{\rm J}(\alpha)$.
The very small $H(\alpha)$ over the range $1.3 \lesssim \alpha \lesssim 2.0$ suggest that these systems have very few ways available to pack more densely, and therefore, in contrast to those discussed in Refs.\ \cite{donev04,delaney05,donev07,vanderwerf18,rocks23,mailman09,schreck10,schreck12}, are nearly maximally stable; note that the maximally-dense monodisperse-ellipse crystal also has $Z = Z_{\rm iso}$.
As $\alpha$ increases past $\sim\! \! 3$, however, the $Z_{\rm J}(\alpha)$ rapidly drop \textit{below} those reported in Refs.\ \cite{delaney05,rocks23}, apparently because employing SWAP and TDOF moves increases the tendency of ellipses to pack into stable  $Z = 4$ configurations where they are trapped by one parallel-aligned neighbor on either side and one unaligned neighbor on either end.
This result is rather surprising because it indicates that maximizing $\phi_{\rm J}$  and maximizing $Z_{\rm J}$  need not always coincide.

\begin{figure*}[!htbp]
\includegraphics[width=6.5in]{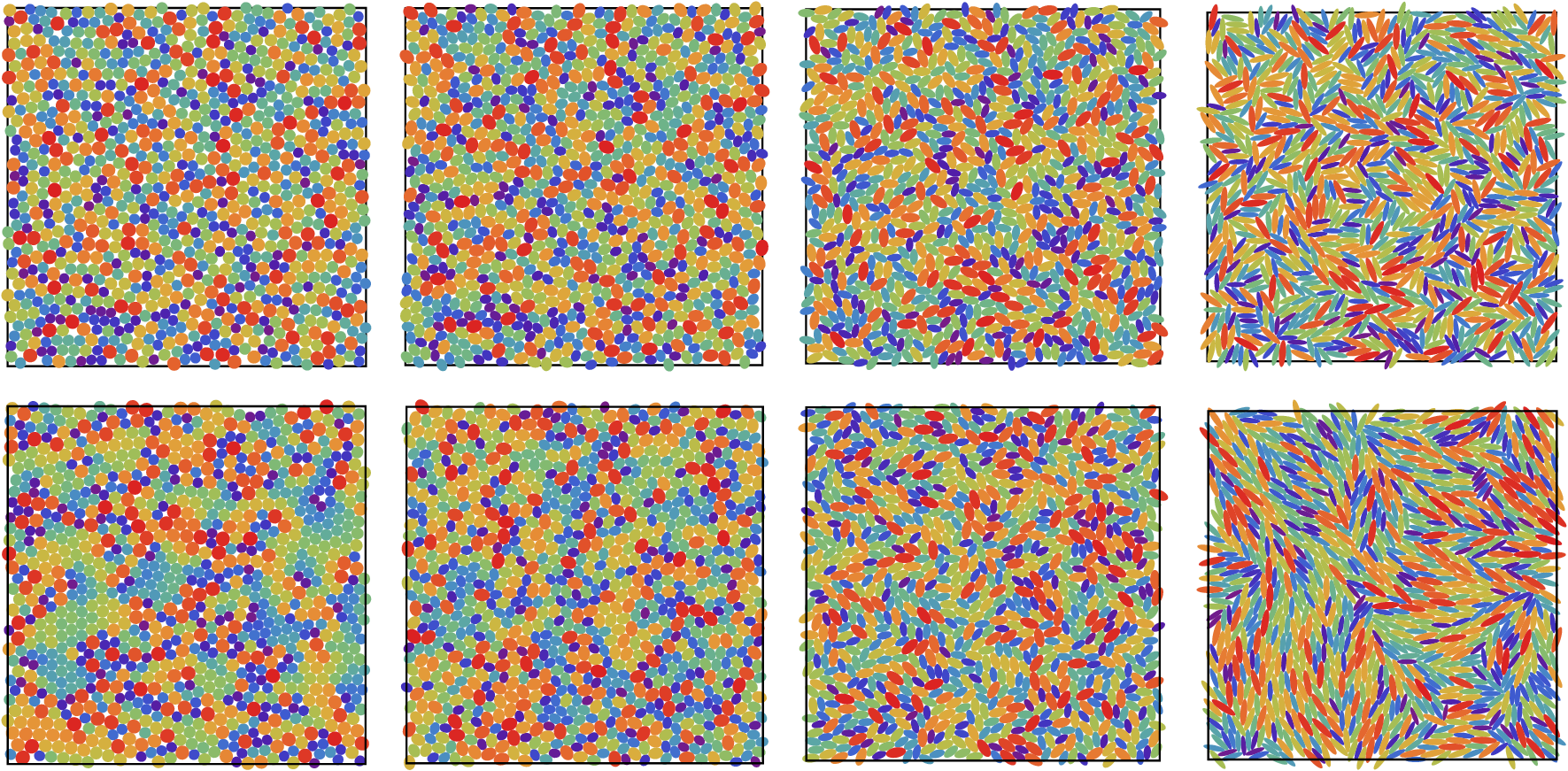}
\caption{Snapshots of typical jammed states with $\alpha = 1.05$, $1.3$, $2$, and $4$ from left to right.  The top (bottom) rows show states prepared without (with) SWAP and TDOF moves.  Particle colors vary from purple to red, in order of increasing $\sigma_i$.}
\label{fig:3}
\end{figure*}

To begin connecting the above results to differences in the packings' multiscale structure, we visually inspected them. 
Typical results for four aspect ratios that illustrate the key trends we observed are shown in Figure \ref{fig:3}.
Results in the top row are similar to those found in previous studies \cite{delaney05,donev07,mailman09,schreck10,schreck12,vanderwerf18,rocks23}.
Those in the bottom row, however, are dramatically different.
For small aspect ratios, adding SWAP and TDOF moves yields strongly fractionated packings consisting of polycrystals of intermediate-size particles, with the largest and smallest particles isolated at the grain boundaries.
The crystalline domains exhibit particle-size gradients whose formation is presumably a collective effect of particle-diameter swapping \cite{ozawa17}.
The grain boundaries contain ``dislocation cores'' which have long been recognized as a distinctive feature of dense polycrystalline disk packings \cite{donev04c}, but have not (to the best of our knowledge) been previously observed in anisotropic-particle packings.

The abovementioned fractionation weakens slowly with increasing $\alpha$, but short-ranged orientational order weakens sufficiently rapidly that the densest packings we obtained ($\alpha = \alpha_{\rm max} = 1.3$) are apparently amorphous despite having a density less than 0.4\% below that of the crystal.  
For $\alpha = 2$, while the packing generated using SWAP and TDOF appears to have greater short-ranged orientational order (to be quantified below), it clearly does \textit{not} include any large locally-nematic domains.
Visual inspection suggests that for these aspect ratios, the packing-efficiency gains achieved  by adding steps (3-4) to the particle-growth procedure appear to be associated primarily with their ability to eliminate most of the sizable voids present in the top-row packings.
We believe that the biased-SWAP moves favor formation of unjammed packings with high $\phi$ and few such voids, and the TDOF moves performed at the end of the packing-generation runs allow formation of extra contacts that bring $Z_{\rm J}$ very close to (i.e.\ within less than 0.5\% of) $Z_{\rm iso}$.

For larger aspect ratios, we find that the increasing packing-efficiency gains highlighted in Fig.\ \ref{fig:1} are directly associated with increasingly-long-ranged orientational order.
Locally-nematic domains are present in the jammed states for $\alpha \gtrsim 3$; their appearance coincides with the beginning of the drops in $Z_{\rm J}(\alpha)$ illustrated in Fig.\ \ref{fig:2}.
In packings generated using SWAP or TDOF moves, these domains look very similar to those found in experimental ``liquid glasses'' formed by ellipsoidal colloids with comparable aspect ratios \cite{zheng11,mishra13,roller20,roller21}.
In packings generated without these moves, the growth of such domains with increasing $\alpha$ is \textit{far} more gradual.
Moreover, an additional distinguishing structural feature is already evident by $\alpha = 4$.
In the top-row (but not the bottom-row) packing, numerous large gaps between differently-ordered domains are visible.
Thus the locally-nematic domains in packings generated using SWAP or TDOF moves, in addition to being larger, \textit{fit together better}, as is evident from the huge reduction in space-wasting tip-to-side contacts visible in this snapshot.

\begin{figure*}[!htbp]
\includegraphics[width=6.5in]{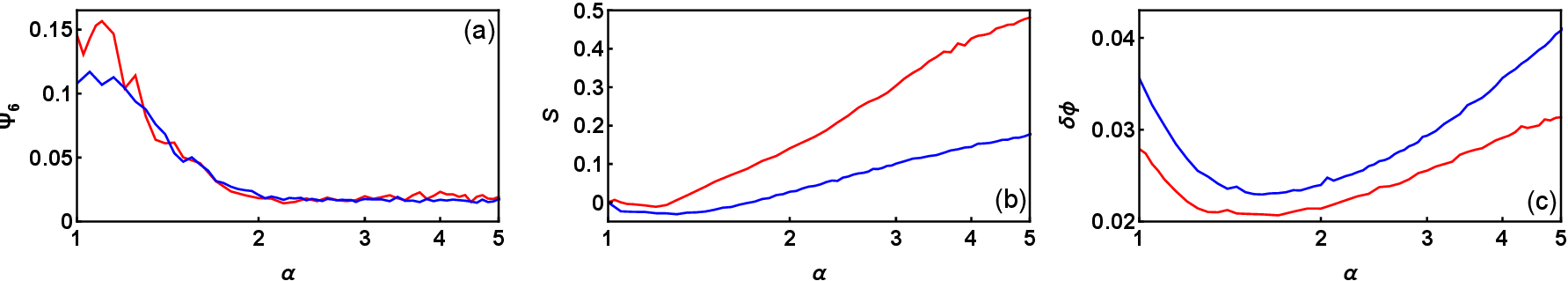}
\caption{ Hexatic order $\Psi_6$ \cite{bernard11}, local nematic order  $S$, and local density fluctuations $\delta \phi$  of systems prepared with and without SWAP and TDOF moves.  All quantities were calculated as described in Ref.\ \cite{rocks23}. Colors are the same as in Figs.\ \ref{fig:1}-\ref{fig:2}.}
\label{fig:4}
\end{figure*}

Figure \ref{fig:4} quantitatively compares the packings' multiscale structure using three additional metrics: the hexatic order parameter $\Psi_6$ \cite{bernard11},  the nematic order parameter $S = \langle [3\cos^2(\Delta \theta_{ij}) -1]/2\rangle$ (where $\Delta \theta_{ij}$ is the orientation-angle difference between ellipses $i$ and $j$), and the density fluctuations $\delta \phi = \sqrt{\langle \phi^2 \rangle - \langle \phi \rangle^2}$.
Here $\Psi_6$ captures orientational ordering on the nearest-neighbor scale, while $S$ snd $\delta \phi$ respectively capture \textit{intermediate-range} orientational and positional order over regions of a size corresponding to a typical particle's first three coordination shells.\footnote[2]{More specifically, as described in Ref.\ \cite{rocks23}, $S$ was calculated using each particles' 18 nearest neighbors, while $\delta \phi$ was calculated using $N$ randomly positioned circular windows of a radius $R$ chosen to make the average window contain 19 particles.}
Since the optimally-dense monodisperse-ellipse crystal with $\phi = \phi_{\rm xtal}$ is simply the triangular lattice affinely stretched by a factor $\alpha$ along one direction \cite{toth50}, it has $\Psi_6(\alpha) = 1 -  \mathcal{O}(\alpha^2)$ for $\alpha - 1 \ll 1$, $S = 1$ for all $\alpha > 1$, and $\alpha$-independent $\delta \phi$.
As might have been expected from the apparent lack of long-range positional or orientational order illustrated in Fig.\ \ref{fig:3}, none of the packings discussed above are close to any of these three limiting behaviors.
On the other hand, Figure \ref{fig:4} also shows that SWAP and TDOF moves strongly affect all three of these structural metrics, and that -- as was the case for $\phi_{\rm J}(\alpha)$ and $Z_{\rm J}(\alpha)$ -- they do so in a strongly-$\alpha$-dependent fashion.

Panel (a) shows that these moves can increase $\Psi_6$ by up to $\sim 50\%$.
This increase is consistent with the formation of fractionated polycrystals discussed above, but it only persists over a narrow range of aspect ratios ($1 \leq \alpha \lesssim 1.15$).
We believe that the sharp drop in $\Psi_6$ over the upper third of this range is responsible for the abovemenentioned minimum in $Z_{\rm J}(\alpha)$ [Fig.\ \ref{fig:2}].

Panel (b) shows that SWAP and TDOF moves increase $S$ over the same range of $\alpha$ for which they increase $\Psi_6$, but only very slightly.
$S$ remains below $.03$ for all $\alpha \lesssim 1.4$, supporting our above claim that the densest packings with $\phi_{\rm J}(\alpha) > .995\phi_{\rm xtal}$ remain amorphous.
On the other hand, adding these moves makes $\partial S/\partial \alpha$ substantially larger for all $\alpha \gtrsim 1.3$.
As long as packings remain effectively isostatic, i.e.\ for $1.3 \lesssim \alpha \lesssim 2$, the resulting differences in $S$ are not associated with the formation of sizable locally-nematic domains.
Instead they appear to be associated with the moves' promotion of side-to-side contacts, which are more space-efficient than tip-to side contacts.
Only for $\alpha \gtrsim 3$, when $S$ exceeds $\sim 0.3$ do such liquid-glass-like domains become apparent (Fig.\ \ref{fig:3}).
Their appearance coincides with the beginning of the rapid increase in packing-efficiency gain and decrease in $Z_{\rm J}(\alpha)$ shown in Figs.\ \ref{fig:1}-\ref{fig:2}.

Panel (c) shows that (i) adding SWAP and TDOF moves substantially reduces $\delta\phi$ for all $\alpha$, and (ii) the fractional reductions in $\delta \phi$ closely track the packing-efficiency gains shown in Fig.\ \ref{fig:1}.
$\delta\phi(\alpha)$ initially decreases with increasing $\alpha$, as the fractionated-polycrystal-plus-dislocation-core structure evident for $\alpha \lesssim 1.15$ gradually gives way to the homogeneous disordered structure evident for $\alpha \simeq \alpha_{\rm max}$.
Its broad minimum, i.e.\ $\delta\phi(\alpha) < .022$ over the range $1.25 \lesssim \alpha \lesssim 2.1$, closely corresponds to the range of aspect ratios over which packings are effectively isostatic (Fig.\ \ref{fig:2}).
For larger aspect ratios, it increases again, but at a slower rate than in packings generated without these moves, consistent with the moves' tendency to make the nematic domains fit together better (Fig.\ \ref{fig:3}).

Finally we briefly discuss the relative contributions of SWAP and TDOF moves to producing the abovementioned differences.
We performed separate runs that omitted growth cycle step (4), and found that the resulting $\phi_{\rm J}(\alpha)$ were only $\sim 0.1\%$ lower, the $Z_{\rm J}(\alpha)$ were substantially lower, the $\Psi_6(\alpha)$ and $S(\alpha)$ did not change significantly, and the $\delta\phi(\alpha)$ were slightly larger.
All trends suggest that the main effect of TDOF moves as employed in \textit{this} study is adding up to $1$ contact per particle at the end of the packing-generation runs.

\section{Discussion and Conclusions}
\label{sec:conclude}

All of the abovementioned structural differences between the ultradense ellipse packings discussed above and those reported in previous studies \cite{delaney05,donev07,mailman09,schreck10,schreck12,vanderwerf18,rocks23} \textit{may} have a single, common explanation.
We hypothesize that they all arise because including biased-SWAP and TDOF moves in the packing-generation procedure allows systems to  escape \textit{kinetic traps} \cite{maher21}.
In other words, including these moves allows systems to bypass the slow dynamics which otherwise lead to jamming at much lower densities.
For low $\alpha$, escaping the traps allow systems to form fractionated polycrystals.
For intermediate $\alpha$, it allows systems to access the slow processes by which small voids are eliminated, and form very-stable isostatic packings,
For large $\alpha$, it allows systems to form much greater local nematic order and shrink the large voids which are otherwise present at the boundaries between differently-oriented domains \cite{rocks23}.
Because the nature of these traps is strongly $\alpha$-dependent, so is the packing-efficiency gain.

Analogous effects have been extensively studied for disk and sphere packings 
 \cite{berthier16,berthier16b,ozawa17,ozawa18,berthier19d,scalliet19,scalliet22,kapteijns19,hagh22,bolton24}, but had not previously been explored for anistropic particles.
Ref.\ \cite{maher21} showed that decreasing the particle growth rate $\tilde{\mathcal{G}}$ in an adaptive-shrinking-cell (ASC)-based algorithm \cite{atkinson12} produces denser, better-ordered packings for a wide variety of particle shapes: rhombi, obtuse scalene and curved triangles, lenses, ``ice cream cones'' and ``bowties.''
It also explained these effects in terms of kinetics, but since it considered only  \textit{monodisperse} systems, did not explore their connection to SWAP or TDOF.
Since employing standard SWAP moves speeds up dynamics by many orders of magnitude in disordered hard-sphere systems above their glass transition densities \cite{berthier16b}, we expect that employing the biased SWAP moves discussed above can be a far more effective method for bypassing anisotropic-particle glasses' kinetic traps than simply decreasing $\tilde{\mathcal{G}}$.

Our results show that all previous studies of polydisperse ellipse jamming \cite{delaney05,donev07,mailman09,schreck10,schreck12,vanderwerf18,rocks23} have failed to access these systems' most-stable disordered jammed states.
The ultradense packings obtained here presumably have vibrational properties which are substantially different from their less-dense counterparts; for example, their far-higher $Z_J(\alpha)$ suggests that they will have far fewer quartic modes \cite{mailman09,schreck10,schreck12}.
Moreover, the effectively-isostatic packings for $\alpha \simeq \alpha_{\rm max}$ may have ideal-glass like vibrational and thermal properties which are the elliptical analogues of those explored in Refs.\ \cite{hagh22,bolton24}.  
Followup studies that employ soft rather than hard ellipses could explore these issues.

Here we have employed a ``maximalist'' biased-SWAP + TDOF approach aimed at generating packings which are as dense as possible while remaining amorphous on large length scales.
However, we emphasize that our method can be generalized to produce packings with any density between those reported in Ref.\ \cite{rocks23} and those reported here, simply by varying the frequency with which the SWAP and TDOF moves are applied.
For example, varying the fraction of particles for which SWAP moves are attempted during step (3), or only performing step (3) periodically, should allow one to systematically study how jammed ellipse packings are affected by sample preparation protocol.
Such studies could improve our understanding of multiple real-world systems composed of anistropic particles whose shapes are sufficiently ellipse-like, including liquid glasses formed by ellipsoidal colloids \cite{zheng11,mishra13,roller20,roller21}, active cell populations \cite{leech24}, and potentially even the trisnapthylbenzenes which have attracted great interest in recent years because some of them form anisotropic quasi-ordered glasses when vapor-deposited \cite{liu15b,liu17b,gujral17,teerakapibal18}.


We dedicate this paper to Mark Ediger for his numerous contributions to our understanding of supercooled liquids and glasses, and thank Madelaine Y.\ Payne for helpful discussions.
This material is based upon work supported by the National Science Foundation under Grant Nos.\ DMR-2026271 and DMR-2419261.



%

\end{document}